\documentclass[twocolumn,prb,aps,english,superscriptaddress]{revtex4}
\usepackage{amsmath}
\usepackage{graphicx}
\usepackage{hyperref}

\newcommand{\av}[1]{\langle{#1}|}
\renewcommand{\v}[1]{|{#1}\rangle}
\newcommand{\bv}[1]{\langle{#1}|}
\renewcommand{\u}{\uparrow}
\renewcommand{\d}{\downarrow}
\newcommand{\U}{\Uparrow}
\newcommand{\D}{\Downarrow}

\begin{document}

\author{Dmitry Solenov}
\author{Sophia E. Economou}
\altaffiliation{sophia.economou@nrl.navy.mil}
\author{T. L. Reinecke}

\affiliation{Naval Research Laboratory, Washington, District of
Columbia 20375, USA}

\title{Fast Two-Qubit Gates in Semiconductor Quantum Dots using a Photonic Microcavity}

\begin{abstract}
Implementations for quantum computing require fast single- and
multi-qubit quantum gate operations. In the case of optically
controlled quantum dot qubits theoretical designs for long-range
two- or multi-qubit operations satisfying all the requirements in
quantum computing are not yet available. We have developed a design
for a fast, long-range two-qubit gate mediated by a photonic
microcavity mode using excited states of the quantum dot-cavity
system that addresses these needs. This design does not require
identical qubits, it is compatible with available optically induced
single qubit operations, and it advances opportunities for scalable
architectures. We show that the gate fidelity can exceed 90\% in
experimentally accessible systems.
\end{abstract}
\maketitle

\section{Introduction}

Quantum information processing involves the manipulation of
entanglement carried out by unitary gate operations between
different quantum bits (qubits). Realistic quantum computing
architectures require entangling gates between distant qubits.
Optical photons provide a natural vehicle to implement such
interactions in many physical systems.\cite{qc} As a result,
architectures based on optically active qubits that can couple to
photonic modes in optical cavities and waveguides, such as quantum
dots, NV centers, and trapped ions are attractive for large scale
quantum computing.
\cite{trappedions,nvctrcavity,nvctrcavity2,reithmeier} Quantum dots
(QDs) in particular hold promise as qubits for such architectures,
in part owing to their large dipole moments, which allow them to
couple efficiently to the optical cavity modes and to photonic
flying qubits for extended architectures. Qubits encoded by the spin
of an electron in a QD have long coherence times which are five to
six orders of magnitude longer than the typical picosecond scale of
optical control. Successful initialization and readout, as well as
fast optical single spin rotations, have been demonstrated in these
systems.\cite{press,spinrot} In addition, important advances have
recently been achieved in work on coupled cavity-QD systems,
including demonstrations of strong coupling and tunability,
\cite{reithmeier,yoshie,badolato,khitrova,hennessy,pinotsi,thon,
gallardo,10to5cavity} and very recently full single-qubit control.\cite{carter_cavity}

A critical step needed to advance the field is the design
of a two-qubit controlled gate operation mediated
by an optical cavity mode. A viable two-qubit quantum gate
requires that several criteria are met: (i) a long-range
switchable physical interaction between qubits is available;
(ii) the gate performs a unitary operation on one
qubit depending on the state of the other qubit to provide
a controlled operation; (iii) the operations are sufficiently
fast compared to decoherence rates; (iv) the gate is compatible with
single-qubit rotations (to form a universal set of gates);
(v) the gate design is consistent with a
multi-qubit system for scalability.

So far, only local control of entanglement in closely spaced quantum
dots (QD `molecules') has been demonstrated
experimentally.\cite{nrlwork} For an experimental demonstration of
cavity-mediated entangling gates, a theoretical design is needed
that satisfies the above criteria, (i)-(v), while being
experimentally simple and compatible with current technology.
Existing proposals for cavity-mediated gates have not met these
requirements; they are either incompatible with single-qubit
gates,\cite{geomphasecavity} limited to nearest-neighbor
qubits,\cite{ladd} and/or require adiabaticity, either through
adiabatic evolution \cite{ladd} or through adiabatic elimination of
the auxiliary state. \cite{imamoglu} As a result, they are much
slower than what is needed from a quantum information processing
perspective. Moreover, a careful assessment of the performance of
such gates as a function of system parameters has not been given in
the literature, despite the key role it would play in experimental
demonstrations.

In this paper we give a novel design for an entangling control-z
(CZ) two-qubit gate \cite{nielsenchuang} that satisfies all the
above criteria. Our design does not require the QD energies to be
equal or dynamically tunable. As a result, our approach is
compatible with single qubit operations and has a potential for
many-qubit scalable architectures. We obtain fidelities in excess of
90\% for realistic parameters. In the following we explain the
concept of this all-optical gate, formulate the model, calculate the
QD-cavity system spectrum, and analyze our design of the two-qubit
gate protocol. The fidelity of the gate operation as a function of
the system parameters is also calculated and provides a guide for
experiment.

\section{Two-qubit gates}

The control-z gate is a maximally entangling two-qubit gate, and it
is given by $ U_{CZ}=\text{diag}(1,1,1,-1)$. It is equivalent to the
more familiar control-NOT (CNOT) operation up to single-qubit gates.
Specifically, $U_{CNOT} = (1 \otimes {\rm H}) U_{CZ} (1
\otimes {\rm H})$, where $H = (\!\!{\tiny \begin{array}{cccc}1 & 1\\
1 & \!\!\!\!-1 \end{array}}\!\!)/\sqrt{2}$ is the Hadamard gate. To
see the entangling capability of \textsc{CZ} we can look at its
action on a product state of two qubits. In particular, when each
qubit is in an equal superposition of the basis states, we have
\begin{eqnarray*}
 U_{CZ}\big(|1\rangle+|0\rangle\big)\otimes\big(|1\rangle+|0\rangle\big)=|11\rangle+|10\rangle+|01\rangle-|00\rangle,
\end{eqnarray*}
which is a maximally entangled two-qubit state, also known as a
two-qubit `cluster state'. Such a state is equivalent to a Bell
state up to single-qubit rotations.

To implement the CZ gate, we need to accumulate a phase factor of
$-1$ selectively to only one of the two-qubit basis states, taken to
be $\v{00}$ above. Meanwhile, to be able to perform single-qubit
gates, the transition involving state $\v{00}$ and an auxiliary
state should be performed in parallel with that involving $\v{01}$
(or $\v{10}$ for rotations of the second qubit) and its
corresponding auxiliary state. To avoid dynamically tuning
energies--a process that is costly in time and can compete with
coherence times--we will use different classes of auxiliary states
for single-qubit and two-qubit operations. In particular, we will
use a near-resonance between the two-photon state of the cavity and
the state where both QDs are excited.

\section{Quantum dots in a cavity}

We focus on a system of two (singly) charged self-assembled InAs QDs
in a photonic crystal microcavity. This structure can support in-
and out-of-plane polarizations.\cite{notomi} Due to strain the
optical dipole transition matrix elements in the InAs dots are
anisotropic, resulting in efficient absorption of light with
electric field polarization perpendicular to the QD growth axis. As
a result, only the mode with electric field polarized in the plane
of the crystal can be coupled to transitions in QDs. We take an
external magnetic field to be applied in-plane (Voigt
configuration), perpendicular to the QD growth direction. This will
enable full single qubit control as explained in
Ref.~\onlinecite{imamoglu}.

\begin{figure}
\centering{}
\includegraphics[width=1\columnwidth]{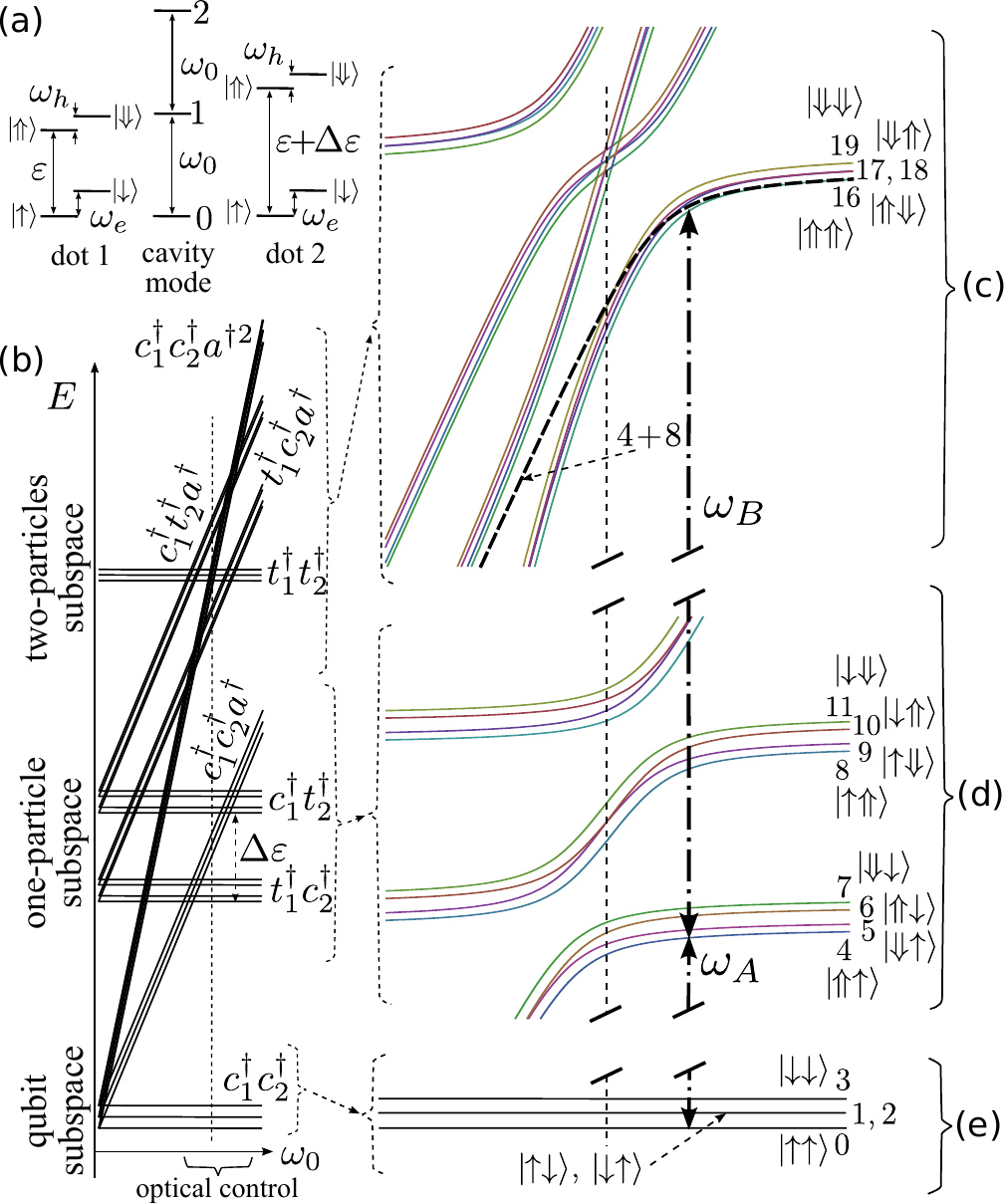}
\caption{\label{fig1}The cavity-dot system. (a) Energies and
relevant states of two QDs and cavity. (b-e) Interacting cavity-dot
spectrum as a function of the cavity mode frequency, $\omega_0$. (b)
Structure of crossings and corresponding states. Panels (c) and (d) show
the anti-crossing splittings in the two- and single-excitation
subspaces respectively. Panel (e) shows the energy structure of the qubit
subspace, which is unaffected by the coupling to the cavity mode.
The numbers in (c-e) give the states of the diagonalized
Hamiltonian, and the $\uparrow$, $\downarrow$ show the predominant
spin character of each state far (to the right) of the avoided
crossings. Vertical dashed lines indicate
$(\varepsilon_1+\varepsilon_2)/2$.}
\end{figure}

The system can be represented by two separate four-state QDs
interacting with a single photon mode, as shown in
Fig.~\ref{fig1}(a). The two lowest energy states of each four-state
system are the spin states of the electron in each dot, which
represent the qubit, $\v{\u}=c^\dag_{n,\u}\v{}_0$ and
$\v{\d}=c^\dag_{n,\d}\v{}_0$, where $n=1,2$ refers to the two dots
and $c^{\dag}_{\u(\d)}$ creates an electron of spin $\u(\d)$
relative to the uncharged QD state $\v{}_0$. The two excited states
in each dot are electron-exciton bound states, called trions (or
charged excitons). They are complexes having total angular momentum
$3/2$. The two $\pm 3/2$ states (`heavy holes') are energetically
lower than the $\pm 1/2$ (`light hole') states and thus form a
pseudo spin $\v{\U}=t^\dag_{n,\u}\v{}_0 =
c^\dag_{n,\u}c^\dag_{n,\d}h^\dag_{n,\U}\v{}_0$ and similarly for
$\v{\D}$, where $h^\dag$ is the creation operator for a heavy hole.
The trion states carry the pseudospin of the hole because the two
electrons are in a spin singlet. We choose the spin quantization
axis along the external magnetic field.

The cavity couples to the trion transitions and preserves the
(pseudo) spin orientation, $\v{\u}\leftrightarrow\v{\U}$ and
$\v{\d}\leftrightarrow\v{\D}$. In the rotating-wave approximation
the cavity-dot interaction is
\begin{equation}\label{eq:H-QC} H_{QD-C} = g\sum_{n=1,2}
\left( t^\dag_{n,\u}c_{n,\u}a + t^\dag_{n,\d}c_{n,\d}a + h.c.
\right)
\end{equation}
where $a$ annihilates a photon in the cavity and $g$ is the coupling
between the trion transition and the cavity. We choose these
coupling constants to be the same for the two QDs to simplify the
presentation. This assumption is not important to the proposed
procedure and can be relaxed when necessary.

The spectrum of the cavity-QD system is shown in Fig.~\ref{fig1} as a
function of the cavity frequency $\omega_0$. This representation
does not suggest the need to tune $\omega_0$ dynamically, but it
helps to identify the region of optimal $\omega_0$ values. The
spectrum is obtained by diagonalizing $H_0 = H_{QD} + H_{C} +
H_{QD-C}$, where $H_{QD} =
\!\!\sum_{n,\xi}\!\omega_e\theta(\xi){c}^\dag_{n,\xi}{c}_{n,\xi}\!\!
+\!\!\sum_{n,\xi}{t}^\dag_{n,\xi}[\varepsilon_n\!\!+\!\omega_h\theta(\xi)]{t}_{n,\xi}$,
$H_{C}=\omega_0a^\dag a$, $\xi=\u,\d$, $\theta(\u)=0$, and
$\theta(\d)=1$.

The Hamiltonian $H_0$ conserves the total number of excitations. As
a result the Hilbert space of the system separates into subspaces
with different numbers of excitations. Each subspace contains
several states, corresponding to different spin projections. The
lowest set of four states (three energy levels) defines the
two-qubit subspace, $\v{\u\u}$, $\v{\u\d}$, $\v{\d\u}$, and
$\v{\d\d}$, and has zero cavity photons; we call this the `zero
excitation' subspace. The other relevant subspaces are the
`one-excitation' subspace, that has states with either one cavity
photon or one trion, and the `two-excitation' subspace, that has
states with two trions (one per dot), states with one trion and one
cavity photon, and states with two cavity photons; see
Appendix~\ref{app:spec}. States in the `one-excitation' part of the
spectrum are approximately local to each quantum dot and interact
with each other only very weakly, $\sim(g/\Delta \varepsilon)^2$.
They are the states that can be used for single-qubit control.
\cite{economouprbprl} The two-excitation subspace involves
hybridized states of the two QDs and are ideal for a two-qubit gate.
These states however are not directly accessible from the qubit
subspace with a single pulse, so we make use of a series of control
pulses.

The laser pulses have momentum perpendicular to the photonic crystal
plane to avoid Bragg shielding due to the photonic crystal. For
definiteness we choose pulses with the same linear polarization as
the cavity mode,\cite{foot1}
\begin{equation}\label{eq:calV}
{\cal V}(t) = \!\!\!\!\!\sum_{p;\,n>m}\!\!\!\!\Omega_p(t-t_p)2\cos\omega_p t\left( M_{nm}\v{n}\bv{m}\! + \!h.c. \right).
\end{equation}
The total Hamiltonian becomes ${\cal H}(t) = {\cal H}_0 + {\cal
V}(t)$, where ${\cal H}_0=U^\dag H_0 U=\sum_nE_n\v{n}\bv{n}$ and
$M_{n,m} = \sum_{j=1,2,\xi}\bv{n}U^\dag(t^\dag_{j,\xi}c_{j,\xi} +
c^\dag_{j,\xi}t_{j,\xi})U\v{m}$. The subscript $p$ enumerates the
pulses used to perform the gate where each has frequency $\omega_p$
and is centered at time $t_p$.

\section{Implementation of CZ gate}

The \textsc{CZ} gate has a simple diagonal form, which allows for a
relatively straightforward design based on phases induced by
resonant cyclic excitation of an auxiliary excited state. The idea
is to use the property of quantum two-level systems, in which a
cyclic evolution from the ground state to the excited state and back
to the ground state induces a minus sign to the latter. In the
presence of additional, uncoupled states the minus sign is relative
and thus constitutes a nontrivial quantum evolution. The pulse
performing such an evolution is known as a `$2\pi$' pulse. Optical
$2\pi$ pulses were proposed theoretically for single-qubit rotations
in quantum dots\cite{economouprbprl} and two-qubit gates in quantum
dot molecules\cite{economou_cphaseQDM} and later used in their
experimental demonstrations.\cite{spinrot,nrlwork}

In our approach, the phase accumulation will be on state $\v{\u\d}$,
while keeping the phases of other basis states unchanged. This can
be done by the following pulse sequence: (i) a population inversion
$\pi$ pulse, pulse A, tuned to transition
$\omega_1=\omega_A=E_4-E_0$ between qubit state $\v{\u\u}$ and the
excited state with similar spin configuration, see
Fig~\ref{fig1}(d)-(e). The pulse is also in resonance with $E_6-E_2$
transition, and thus it creates a trion in the first QD only: both
$\v{\u\u}$ and $\v{\u\d}$ are transformed in the same way and
accumulate a phase factor of $-i$ each. (ii) A $2\pi$, or phase,
pulse (pulse B) with frequency $\omega_2=\omega_B=E_{16}-E_4$, see
Fig~\ref{fig1}(c)-(d). This induces a transition between the
`one-excitation' states previously created and one of the
`two-excitation' states. Note that if $g=0$ or we are far detuned,
$\omega\gg \Delta\varepsilon$, the transition $E_{10}-E_2$ would
also occur. This would correspond to a single qubit operation on the
second qubit, i.e., $\v{\U\u}$ and $\v{\d\u}$ would both acquire a
phase factor of $-1$. A nonzero $g$ induces formation of
two-excitation states that have different energies, c.f. the energy
of state 16 and the sum of energies of states 4 and 8. As a result,
the state $\U\u$, or state 4, acquires the factor of $-1$ after the
pulse, while state $\v{\d\u}$ does not. (iii) Finally, we apply the population
inversion pulse A again, $\omega_3=\omega_A$, to restore the system
to the qubit subspace. This gives additional factors of $-i$ to both
$\v{\u\u}$ and $\v{\u\d}$, as mentioned above. The two phase factors
of $(-i)$ induce a minus sign to states $\v{\u\d}$ and $\v{\u\u}$,
while the $2\pi$ pulse cancels that sign in state $\v{\u\u}$. The
phase between the control pulses A and B does not enter the result
and therefore pulses with unequal frequencies do not have to be
phase locked, which is a significant experimental convenience.

A physical explanation of this approach is the following: because
each QD is off-resonant from the cavity, when only one of the QDs is
excited and no other excitations are present in the system the
excited QD can be roughly thought of as isolated, i.e., decoupled
from the cavity and from the other QD. Thus, single excitations can
implement single-qubit operations without disturbing the rest of the
system. On the other hand, when both QDs are excited they are closer
to the resonance with the cavity state. As a result, there is a
large mixing between cavity states and the states of both QDs. Thus,
using the two-excitation regime is a natural venue for performing
two-qubit conditional operations while maintaining the ability to
manipulate each QD spin separately.

\section{Fidelity}

Now we consider the gate fidelity, which is a measure of how close
our operation is to the target gate. There are two types of fidelity
losses, those caused by unintended coherent dynamics due to coupling
of the lasers to off-resonance transitions and those originating from
random processes such as trion recombination. First we focus on the
former mechanism. The unintended transitions can cause $U_g$ to
deviate from the ideal $U_{CZ}$ and effectively cause loss of
coherence in the qubit-subspace, even though the entire operation
involving excited states is unitary and coherent.

\begin{figure*}
\includegraphics[width=1.3\columnwidth]{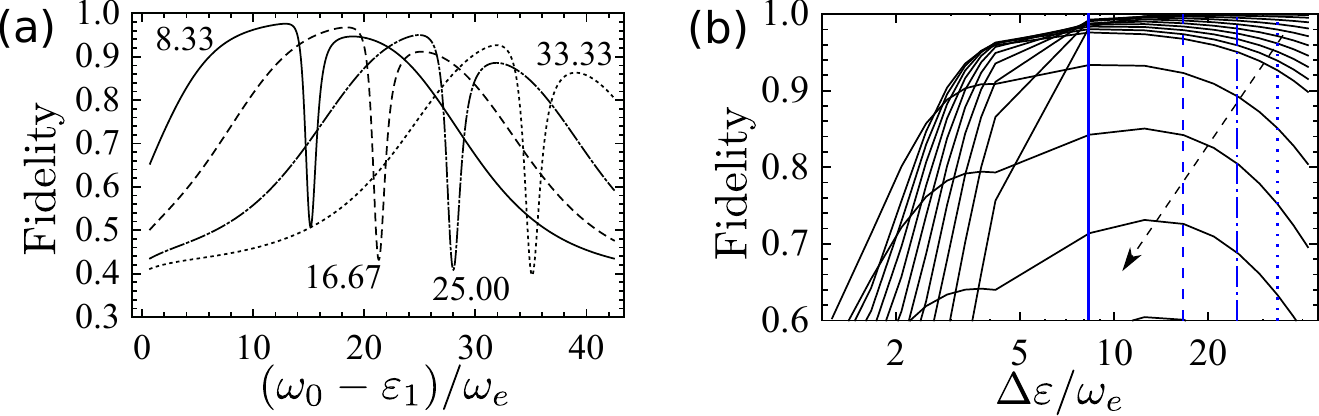}
\caption{\label{fig2-1}Fidelity of CZ gate with imperfections
resulting from coupling of the pulses to neighboring off-resonance
optical transitions (a) as a function of $\omega_0$ for
$\sigma/\omega_e=0.1$ for $\Delta\varepsilon/\omega_e = 8.33$,
16.67, 25.00, 33.33, as indicated, and (b) as a function of the
spectral separation $\Delta\varepsilon$ between the QDs for
different values of the pulse bandwidth, $\sigma/\omega_e = 0.01$,
0.02, ..., 0.1, 0.15, ..., 0.3 as indicated by the dashed arrow.
Each point is computed for the optimal value of $\omega_0$ from Fig.
\ref{fig2-1}(a). The vertical lines mark the values of $\omega_2$ from
panel (a). In both panels (a) and (b) we used $g/\omega_e=3.33$ and
$\omega_h=\omega_e/3$.}
\end{figure*}

\begin{figure*}
\includegraphics[width=1.4\columnwidth]{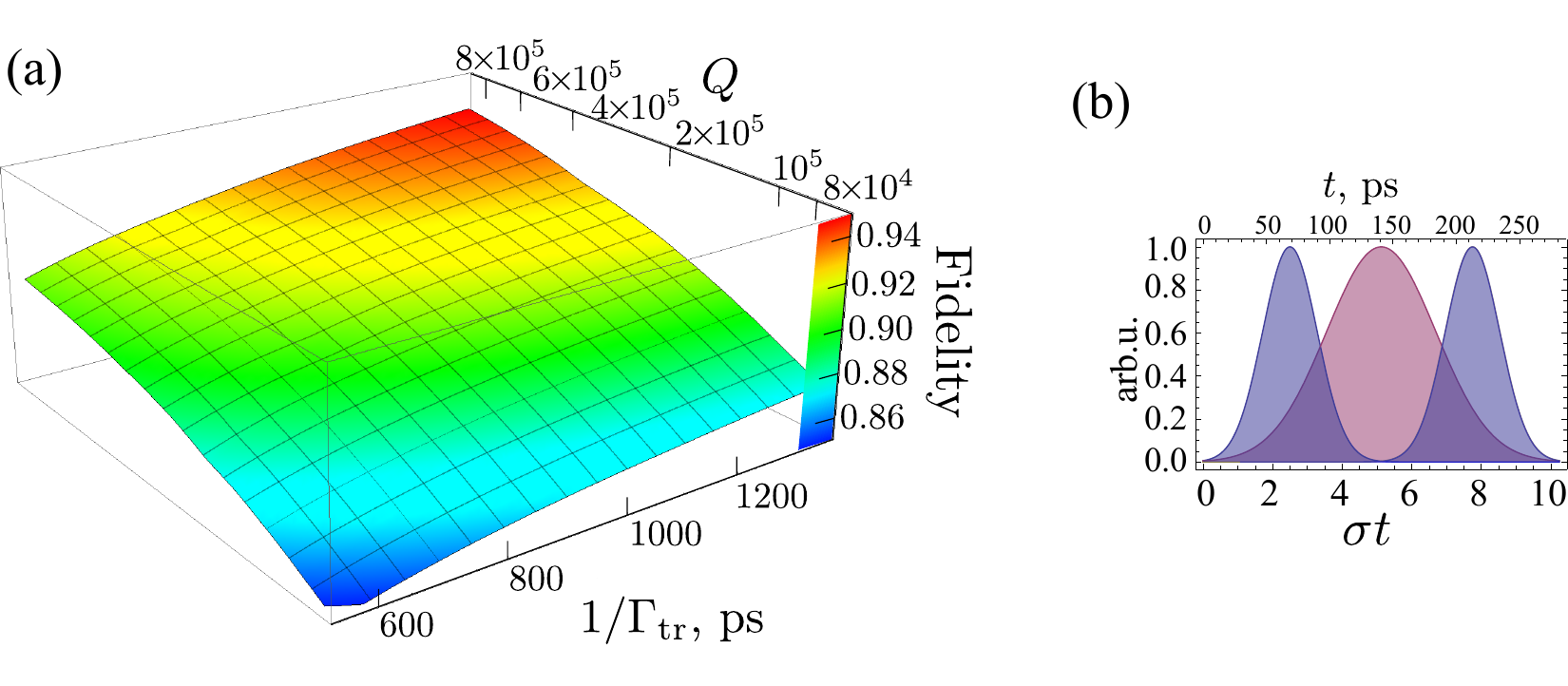}
\caption{\label{fig2-2}(a) Fidelity of the two-qubit CZ gate in
presence of decoherence due to trion recombination and cavity decay.
The fidelity is plotted as the function of the trion decay time and
the cavity mode quality factor. (b) The temporal profiles of the
pulse sequence for $\sigma/\omega_e = 0.2$, $\omega_h=\omega_e/3$,
$\Delta\varepsilon/\omega_e=8.33$, $g/\omega_e=3.33$, and
$\omega_e=0.12$meV.}
\end{figure*}

To analyze this type of decoherence we compute the average fidelity,
$F$, of the gate operation, as explained in detail in
Appendix~\ref{app:fidel}, including transitions 0-4, 4-6, 2-5, 3-7,
0-8, 2-10, 1-9, 3-11 for pulse $A$ and transitions 4-16, 6-18, 0-8,
2-10, 1-9, 3-11 for pulse B. Other transitions are negligible either
due to vanishing matrix elements or to large detuning. We chose
different pulse widths for pulse A and pulse B $\sigma_A = 2\sigma_B
= 2\sigma$. In Fig.~\ref{fig2-1}(a) the fidelity is plotted as a
function of the difference between the cavity mode frequency
$\omega_0$ and the transition frequency of QD1 $\varepsilon_1$ for
varying values of the frequency of QD2 $\varepsilon_2$. The
qualitative features of the plots can be understood as follows: when
the cavity mode frequency is much smaller or much larger than the QD
frequencies, QD-cavity hybridization is negligible, and we are in a
regime of two independent qubits. This causes attenuation of
fidelity towards both sides of the plot. The dip in the middle
occurs because, as the cavity is tuned, the target transition of
pulse B (transition 4-16) becomes degenerate with transition 3-11,
and therefore state $|\downarrow\downarrow\rangle$ is also affected
by pulse B, resulting in strong unintended dynamics. Note that at
its high values the fidelity does not vary strongly with
$\varepsilon_1$ and $\varepsilon_2$. As a result, gates between
several different pairs of quantum dot spin qubits can be performed
with high fidelity using only one cavity mode to mediate the
interactions, which is an intriguing opportunity for scalable
architectures.

Fig.~\ref{fig2-1}(b) shows the fidelity as a function of the spectral
separation $\Delta\varepsilon$ between the trions in QD1 and in QD2
for different pulse bandwidths $\sigma$. When $\Delta\varepsilon$ is
small (comparable to $\omega_e$) the fidelity drops appreciably.
This drop is the result  of coupling in the `one-excitation'
subspace, i.e., the assumption that an excited QD is isolated from
the rest of the system is no longer valid. Thus it also identifies
the regime where fast optical single-qubit control is not possible.
In the region of higher fidelities, where
$\Delta\varepsilon/\omega_e\gtrsim 10$, the fidelity approaches its
maximal value for longer pulses and starts decreasing more rapidly
for $\sigma/\omega_e\gtrsim 0.2$ due to involvement of a larger
number of unintended transitions.

Next, we consider the effects of decoherence due to losses during
the gate. The main contributions come from trion recombination and
cavity photon leakage. The typical linewidth of the trion state,
$\Gamma_{\rm tr}$, in InAs QDs is $\sim 1\mu$eV. \cite{cortez} The
loss rate associated with the cavity is $\Gamma_c = \omega_0/Q$.
State-of-the-art microcavities \cite{10to5cavity} can have $Q$'s up
to $\sim 10^5$, which gives $\Gamma_c\sim 10\mu$eV. We calculate the
fidelity using the standard master equation formalism
\cite{nielsenchuang,solenov} and include states from 0 to 19, see
Appendix~\ref{app:fidel}. The fidelity as a function of $Q$ and
$1/\Gamma_{\rm tr}$ in shown in Fig.~\ref{fig2-2}. It is maximized
when the pulses overlap to reduce the time the excited states are
occupied. We choose $\omega_0$ from the maximal fidelities, as in
Fig.~\ref{fig2-1}(a) for each point of Fig.~\ref{fig2-2}. We see
that fidelities in excess of 90\% are possible for realistic values
of the parameters.

\section{conclusions}
In summary, we have developed a design for a cavity-mediated
entangling gate between two spin qubits that satisfies the criteria
for a realistic two-qubit operation. Our control-z gate is
compatible with available single qubit operations and with natural
inhomogeneities in optical resonances. It can thus accommodate
several qubits that couple pairwise with appropriate control laser
frequencies, opening a path to scalable architectures. It may also
be useful for hybrid quantum computing approaches with various
physical systems.\cite{hybrid} We have shown that the gate fidelity
is at least 90\% for current experimental parameters. Higher
fidelities can be achieved in various ways such as using pulse
shaping techniques \cite{oct,economoupulses} and engineering higher
finesse cavities.

\section{acknowledgements}
This work was supported in part by NSA/LPS and in part
by ONR.

\appendix

\section{The spectrum}\label{app:spec}

In the rotating wave approximation the Hamiltonian [Eq. (1) from
the main text] conserves the total number of excitations. Therefore it
can be diagonalized independently in each excitation-number
subspace. The lowest energy set of four states (three energy levels)
corresponds to a subset with zero excitations.  It represents the
two-qubit subspace with zero cavity photons,
\begin{eqnarray}\label{eq:qsubspace}
0\to\v{\u\u}|0\rangle,
1\to\v{\u\d}|0\rangle,
2\to\v{\d\u}|0\rangle,
3\to\v{\d\d}|0\rangle,
\end{eqnarray}
where $|0\rangle$ is the vacuum state of the cavity. The
corresponding energies are controlled by the  magnetic field via
Zeeman splitting. For typical values of magnetic field of $\sim 1$ T
used in the initialization and readout and single-qubit experiments
the splitting between $E_0$,$E_3$ and $E_{1,2}$ is $\sim 0.1$ meV.
The micro-cavity optical mode is coupled to the excitonic
transitions in each quantum dot with the transition energies $\sim$
eV. As a result, the qubit subspace is not affected by the cavity.

The one-excitation subspace occurs at the optical frequency, $\sim$
eV from  the qubit subspace energies:
\begin{eqnarray}
\label{eq:1space1}
\text{dot 1}&:&\v{\U\u}|0\rangle,\v{\D\u}|0\rangle,\v{\U\d}|0\rangle,\v{\D\d}|0\rangle,\\
\label{eq:1space2}
\text{dot 2}&:&\v{\u\U}|0\rangle,\v{\u\D}|0\rangle,\v{\d\U}|0\rangle,\v{\d\D}|0\rangle,\\
\label{eq:1spaceC}
\text{cavity}&:&\v{\u\u}\v{1},\v{\u\d}\v{1},\v{\d\u}\v{1},\v{\d\d}\v{1},
\end{eqnarray}
where $\v{1}$ denotes the state with a single photon in the cavity.
The energy gap $\Delta\varepsilon$  between states
(\ref{eq:1space1}) and (\ref{eq:1space2}) is due to the fact that
the two dots are not identical in size, shape, and strain
environment, which affects the excitonic transitions. The typical
variation in trion transition energies is $\sim 1-20$ meV. The
energy of the cavity mode, $\omega_0$, is fixed during the gate
operation but can be set to an optimal value during sample growth.
The interaction with a cavity photon shifts the energies and mixes
trion and photon states. The energies of the resulting states can be
found analytically: note that states
(\ref{eq:1space1}-\ref{eq:1spaceC}) are always coupled in triplets.
For example, the state $\v{\u\u}\v{1}$ interacts only with
$\v{\U\u}\v{0}$ and $\v{\u\U}\v{0}$. For each triplet we have
\begin{eqnarray}\label{eq:spec3}
(E\!\!-\!\varepsilon_{1,\xi})(E\!\!-\!\varepsilon_{2,\xi})(E\!\!-\!\omega_0)\!=\!g^2\!(E\!\!-\!\varepsilon_{1,\xi})\!\!+\!g^2\!(E\!\!-\!\varepsilon_{2,\xi})\!,
\end{eqnarray}
where $\xi = \u$ or $\d$, $\varepsilon_{n,\u}=\varepsilon_n$ and
$\varepsilon_{n,\d}=\varepsilon_n+\omega_h-\omega_e$. Each triplet
forms two anti-crossings when $\omega_0$ is swapped across the trion
energies  [see Fig.~1(b) and Fig.~1(d) of the main text]. When
$g\sim\Delta\varepsilon$ or $g\gg\Delta\varepsilon$, the two excited
quantum dot states can mix and form spin-entangled states. For
experimentally accessible systems of quantum dots in a micro-cavity
the coupling strength $g$ is substantially smaller than the
variation in trion energies $\Delta\varepsilon$ and the mixing is
negligible. In the limit $g\ll\Delta\varepsilon$ the interaction
between one-excitation states from different QDs can be estimated by
analyzing the difference $\delta\omega_\u$ in transition energies
between $\omega_\u:\v{\u\u}\to\v{\U\u}$ and
$\omega'_\u:\v{\u\d}\to\v{\U\d}$. From Eq.~(\ref{eq:spec3}) we find
$\omega_\u=\varepsilon+g^2/f(\omega_A,\Delta\varepsilon)$ and
$\omega'_\u=\varepsilon+g^2/f(\omega_A',\Delta\varepsilon+\omega_e-\omega_h)$,
where $f(y,x)=x-\omega_0-g^2/(x-\varepsilon+y)$. Since
$\omega_e\sim\omega_h\sim g\ll\Delta\varepsilon$ it is easy to show
that $\delta\omega_\u \lesssim - g^2\omega_e/\Delta\varepsilon^2$.
This should be compared to the typical inverse lifetime of the trion
state, $\sim 1 \mu$eV (in energy units) or $\sim\omega_e/100$. As a
result for $\omega_e/\Delta\varepsilon\sim 10$, $\omega_\u$ and
$\omega'_\u$ are practically indistinguishable. This result is
confirmed numerically by computing the spectrum (and the states) for
different values of $\Delta\varepsilon$. It also holds for other
transitions between the qubit and the one-excitation subspace
states. Therefore we conclude that the one excitation subspace
cannot be used for a two-qubit operations.  It can, however, be used
to perform fast single qubit operations as described in
Ref.~\onlinecite{economouprbprl} by using the localized trion state.

In order to find useful non-local states that can mediate a
two-qubit gate  we investigate the two-excitation subspace. In this
subspace the states are coupled in groups of four, e.g.
$\v{\U\U,0}$, $\v{\U\u,1}$, $\v{\u\U,1}$, $\v{\u\u,2}$:
\begin{eqnarray}\label{eq:spec4}
(\varepsilon_{2,\xi}\!+\!\varepsilon_{1,\xi}\!-\!E)(\omega\!+\!\varepsilon_{1,\xi}\!-\!E) (\omega_0\!+\!\varepsilon_{2,\xi}\!-\!E)(2\omega\!-\!E)
\\\nonumber
= g^2(\varepsilon_{2,\xi}\!+\!\varepsilon_{1,\xi}\!+\!2\omega_0\!-\!2E)^2.
\end{eqnarray}
The spectrum has a more complex structure, see Fig.~1(c) of the main
text. The two-excitations subspace provides non-local
quantum-dot-cavity states, such as state 16, which has two trions
(one in each dot). The energy of such state is different from the
combined energy of two trion states localized in each dot, such as
$E_{4}$ and $E_{8}$,
\begin{eqnarray}\label{eq:onevstwo}
\Delta E_{16,4} \neq \Delta E_{4,0} + \Delta E_{8,0}
\end{eqnarray}
where $\Delta E_{n,n'} = E_n - E_{n'}$. This is the basis for the
two-qubit  conditional phase gate in this work. Using a perturbative
approach like that above, we obtain $\Delta E_{16,4}-(\Delta E_{4,0}
+ \Delta E_{8,0}) \sim -g^2/\Delta\varepsilon$.

\section{Gate Fidelity}\label{app:fidel}

The fidelity of the gate described in the main text is affected by
two type of processes: (i) induced unintended transitions between
the states of the qubit-cavity system and (ii) real losses due to
cavity leakage and trion recombination. We first estimate losses due
to unintended but coherent dynamics. We include transitions 0-4,
4-6, 2-5, 3-7, 0-8, 2-10, 1-9, 3-11 for pulse $A$, and 4-16, 6-18,
0-8, 2-10, 1-9, 3-11 for pulse B ($\omega_B$). Other transitions are
negligible either due to vanishing matrix elements or to large
detuning. We compute the wave function after the A-B-A pulse
sequence for each basis configuration of the qubit subspace as
initial state (evolution is linear and therefore the resultant wave
function for any initial qubit state can be easily recovered). To
simplify calculations here we resort to analytically solvable
Rosen-Zener pulse  shapes,\cite{rosenzener} i.e.
$\Omega_p(t)=\Omega_p{\rm sech}(\sigma_p t)$ with $\sigma_A =
2\sigma_B = 2\sigma$, to calculated transition amplitudes and phases
for resonant and off-resonance transitions for each pulse. Given the
initial, $\v{\psi_0}$, and final, $\v{\psi} = U\v{\psi_0}$, wave
function, the fidelity can be computed as
\begin{eqnarray}\label{eq:coherentF}
F(\psi_0,\psi) = |\bv{\psi_0}U^\dag_{CZ}\v{\psi}|
\end{eqnarray}
where $U^\dag_{CZ}$ is the evolution operator corresponding to the
ideal CZ gate. The value of $F(\psi_0,\psi)$ depends on the initial
state of the two-qubit system and therefore can vary depending on the
choice of algorithm and initial data. We therefore compute the
average fidelity $F$ by taking average over all possible initial
states of the two-qubit system,
\begin{eqnarray}\label{eq:coherentAvF}
F^2&=& \int d\psi_0 F(\psi_0,\psi\{\psi_0\})^2
\\\nonumber
&=&
\sum_{ijnm}
\frac{\delta_{in}\delta_{jm} + \delta_{ij}\delta_{nm}}{20}
\bv{n}U^\dag_{CZ}U\v{i}\bv{j}U^\dag U_{CZ}\v{m}
\end{eqnarray}
The integration $\int d\psi_0$ is performed over all complex
amplitudes that define the initial state in the basis $\v{i}$, and
$i,j,n,m$ run over all basis states $\u\u,\u\d,\d\u,\d\d$.
\cite{fidel} The results are presented in Fig.~2 and the discussion
is given in the main text.

In order to account for both unintended dynamics and actual losses
we have to calculate the reduced density matrix, $\rho(t)$, of the
two-qubit sub-system for the duration of the pulse sequence. The
reduced density matrix can be found within the Bloch-Redfield
master-equation (ME) formalism
\begin{eqnarray}\label{eq:rho-eq}
&&i\dot{\rho} = [H+V(t),\rho]
\\\nonumber
&&+\sum_s
i\Gamma_s\left[P_s\rho P_s^\dag - \frac{P^\dag_sP_s\rho+\rho P^\dag_s P_s}{2}\right]
\end{eqnarray}
where $P_s = \v{f_s}\bv{i_s}$, $\v{i_s}$ and $\v{f_s}$ are initial
and finial states (in the spin basis) corresponding to the $s$-th
decay process with rate $\Gamma_s$. Solving the above equation
directly is computationally involving due to the presence of two
different time scales: fast, associated with the laser driving
frequency, and slow, coming from the time-dependence of the pulse
shaped envelope. To simplify the computation we transform the ME to
the eigenbasis of $H$ and use the rotating wave approximation,
\begin{eqnarray}\label{eq:rhoRW}
&&\dot{\tilde\rho} = -i[{\frak V}(t),\tilde\rho]
\\\nonumber
&&+\sum_s
\Gamma_s\left[{\cal P}_s\tilde\rho{\cal P}_s^\dag - \frac{{\cal
P}^\dag_s{\cal P}_s\tilde\rho+\tilde\rho{\cal P}^\dag_s{\cal
P}_s}{2}\right],
\end{eqnarray}
where $\tilde\rho = e^{i{\cal H}_0t}U^\dag\rho U e^{-i{\cal H}_0t}$, ${\cal P}_s = U^\dag P_s U$ and
${\frak V}(t)=e^{i{\cal H}_0t}{\cal V}(t)e^{-i{\cal H}_0t}$.
Note that the trion decay processes can involve photons with any in-plane polarization (along or
perpendicular to the applied magnetic field). Therefore, for each
trion we have $\Gamma_s=\Gamma_{\rm tr}$: $P_s\to\{\v{\u}\av{\U},
\v{\d}\av{\U}, \v{\u}\av{\D}, \v{\d}\av{\D}\}$. Leakage of photons
from the cavity is modeled as $\Gamma_s=\Gamma_c$:
$P_s\to\{\v{0}\av{1}, \sqrt{2}\v{1}\av{2}, \text{etc}.\}$.
Due to additional (pseudo)spin-flip electron-hole recombination processes,
more states are involved than for the coherent case discussed
above and we include states from 0 to 19. We chose to use Gaussian pulse
shapes $\Omega_p(t)=(\Omega_p/\sqrt{\pi/2})
\exp\{-2t^2\sigma_p^2/\pi^2\}$ for numerical convenience and apply
the same pulse sequence as before with $\sigma_A = 2\sigma_B = 2\sigma$.

Since a separable quantum wave function is no longer accessible, fidelity has to be defined differently,
\begin{eqnarray}\label{eq:rhoF}
F(\psi_0,\rho\{\psi_0\}) = \sqrt{\bv{\psi_0}U^\dag_{CZ}\,\rho\, U_{CZ}\v{\psi_0}}
\end{eqnarray}
In this case the average fidelity is computed as
\begin{eqnarray}\label{eq:rhoF}
F^2\!\!\! = \!\!\!\!\!\!\!\!\!\!\!\sum_{ijnm=\{1,4\}}
\!\!\!\!\!\!\!\!\!\frac{\delta_{in}\delta_{jm} + \delta_{ij}\delta_{nm}}{20} \bv{n}U^\dag_{CZ}\rho\{\v{i}\bv{j}\}U_{CZ}\v{m}
\end{eqnarray}
which is the generalization of Eq.~(\ref{eq:coherentAvF}) for the
case of non-unitary evolution of pure initial state. This is
possible due to the fact that the evolution of the density matrix is
still described by a linear (but non-unitary) super-operator, i.e.
$\rho(t) = T\exp(-i\int_0^tdt L_H(t)-tL)\v{\psi_0}\bv{\psi_0}$,
where $L_H O = [H,O]$. As a result, the complex coefficients that
define initial ($\v{\psi_0}$) and target ($U_{CZ}\v{\psi_0}$) states
in the basis $\v{i}$ can be integrated out in exactly the same way
as for Eq.~(\ref{eq:coherentAvF}). The results are presented in
Figs.~2 and 3 in the main text.

\end{document}